\documentclass[twocolumn,10pt]{article}
\usepackage{aas}
\usepackage{bm,upgreek}
\title{Modeling the meteoroid environment far from the ecliptic plane}

\author[1]{Althea V.\ Moorhead}
\affil[1]{NASA Meteoroid Environment Office, Marshall Space Flight Center, Huntsville, AL 35812, USA; \href{althea.moorhead@nasa.gov}{althea.moorhead@nasa.gov}}

\author[2,3]{Petr Pokorn\'y}
\affil[2]{Astrophysics Science Divison, NASA Goddard Spaceflight Center, Greenbelt, MD, 20771, USA}
\affil[3]{Department of Physics, The Catholic University of America, Washington, DC, 20064, USA}

\author[4]{Marcus A.\ Holden}
\affil[4]{Universities Space Research Association, 320 Sparkman Drive Huntsville, AL 35805 USA}

\author[5]{William Kosmann}
\affil[5]{The Astronautics Company, L.P., 37523 Maple Shade Lane Middleburg, VA 20117 USA}

\begin{document}

\section{Introduction}
\label{sec:intro}

NASA's Meteoroid Engineering Model (MEM) describes the flux, directionality, speeds, and densities of hazardous meteoroids in the inner solar system \citep{moorhead20}. The model uses a physics-based approach to extrapolating this information from meteoroid observations collected near Earth. The farther removed locations are from Earth, the more uncertain this extrapolation is; for this reason, the model generates environment descriptions only for spacecraft trajectories that remain between 0.2 and 2~au from the Sun.

MEM has also historically assumed that the spacecraft encountering the environment lies in or close to the ecliptic plane (i.e., the plane encompassing the Sun and Earth's orbit). This is a perfectly reasonable assumption for the vast majority of missions: any spacecraft orbiting Earth, for instance, will necessarily remain close to the ecliptic. However, it is possible for a spacecraft's trajectory to take it far from the ecliptic: \emph{Ulysses} used a gravity assist from Jupiter to pass over the Sun's polar regions with an ecliptic latitude of $80.2^\circ$. We have advised users that MEM~3 is not valid at these ecliptic latitudes \citep{memioc}.

With a heliocentric orbital inclination of $80^\circ$, \emph{Ulysses} has reached higher ecliptic latitudes than any other spacecraft. However, Solar Orbiter will reach an orbital inclination of $24^\circ$ by the end of its primary mission, and may reach an inclination of $33^\circ$ during an extended mission \citep{muller20}. Trajectories with high ecliptic latitudes continue to be proposed for solar observation missions \citep{sekii15}, one recent example of which is \emph{Solaris} \citep{hassler23}. A modified version of MEM is needed in order to accurately compute the risk of meteoroid impacts along such a trajectory. Such a version of MEM will also be useful for validating meteoroid models against impacts on asteroids \citep[see, e.g.,][]{bottke20}, which do not necessarily have low orbital inclinations.

All versions of MEM assume that the secondary orbit angles (argument of perihelion, longitude of ascending node, and mean anomaly) of sporadic meteoroids have been fully randomized, and all previous versions of MEM assume that the observer or spacecraft lies in the ecliptic plane. Under these assumptions, the meteoroid number density is proportional to the sum of the probabilities of collision between each modeled meteoroid orbit and a second, non-inclined orbit \citep{opik51}. With MEM~3.1, we remove the assumption that the observer lies in the ecliptic and calculate the spatial density for arbitrary ecliptic separation \citep{kessler81,steel85}. We adopt the approach recommended in Appendix~A of \cite{kessler81} for calculating finite spatial densities at certain locations; however, we find that it is computationally feasible to use this approach at all locations. We calculate the encounter geometry and speed between each modeled meteoroid orbit and an observer using an approach equivalent to that of \cite{divine93}. These algorithms are described in Section~\ref{sec:math}; they have also been implemented in Python and made available through NASA's public GitHub \citep{repo}.

We find that our updated algorithms result in a vastly reduced meteoroid flux at high ecliptic latitudes. For instance, the meteoroid flux on \emph{Solaris} when it is at high latitudes is approximately 20\% the flux it would encounter if it remained near the ecliptic (see Sec.\,\ref{sec:solaris}). However, when we compare the distribution of zodiacal light corresponding to MEM's meteoroid orbital distributions, we find that it is more concentrated near the ecliptic than the IRAS data \citep{nesvorny10}. Further modeling of meteoroid orbits will be needed to bring MEM's orbital element distributions into agreement with observations.

\section{Number density and directionality}
\label{sec:math}

The key advantage that MEM~3.1 offers over its predecessors is that MEM~3.1 correctly computes the flux and directionality of meteoroids relative to spacecraft trajectories that venture far from the ecliptic plane. 

When creating MEM, \cite{jones04} made the following assumptions: [1] the observer or spacecraft lies in the ecliptic plane; and [2] some of the meteoroids' orbital elements -- argument of pericenter, $\omega$; longitude of ascending node, $\Omega$; and mean anomaly, $M$ -- are randomized and follow a uniform distribution, ${U(0, 2 \pi)}$. These are reasonable choices when modeling the near-Earth meteoroid environment. However, the first assumption is not valid for an observer or spacecraft that ventures far from the ecliptic plane. In this section, we present an approach for modeling the meteoroid environment that removes this assumption in order to handle high ecliptic latitudes.

We will, however, retain the second assumption of \cite{jones04} that the three orbital angles $\omega$, $\Omega$, and $M$ have been randomized. This assumption allows us to force a meteoroid with nearly any given combination of semi-major axis, $a$; eccentricity, $e$; and inclination, $i$, to intersect a given spacecraft location. We also relax the requirement that the spacecraft must lie between the meteoroid's perihelion and aphelion distances (${q \le r \le Q}$) and also closer to the ecliptic plane (${\sin \beta \le \sin i}$); instead, we will require that the spacecraft lie within a certain characteristic distance ($d$) of these boundaries.

A simple, open-source implementation of our number density and encounter geometry algorithms is available online \citep{repo}.

\subsection{Spatial distribution}
\label{sec:pdfs}

The heliocentric distance ($r$) and ecliptic latitude ($\beta$) of our target location can be converted to the following unitless variables:
\begin{align}
    s &= (r-a)/ae \label{eq:s} \\
    \xi &= \sin \beta / \sin i \label{eq:xi}
\end{align}
Both variables lie within the interval ${[-1, 1]}$, and their joint probability distribution within that range is:
\begin{align}
    f_{s \xi}(s, \xi) &= f_s(s) \cdot f_\xi(\xi) 
        \, \text{, where } \label{eq:fsx} \\
    f_s(s) &= \frac{1}{\pi} \frac{1 + e s}{\sqrt{1 - s^2}} 
        \, \text{ and } \\
    f_\xi(\xi) &= \frac{1}{\pi} \frac{1}{\sqrt{1 - \xi^2}}
        \, .
\end{align}
A derivation of the above equations is presented in appendix\,\ref{sec:derivpdf}. These equations are equivalent to eq.~21 of \cite{kessler81} and eq.~25 of \cite{steel85}; this is demonstrated in appendix\,\ref{sec:equiv}. Note that $f_{s \xi}$ is non-finite when ${s = \pm 1}$ or ${\xi = \pm 1}$.

\begin{figure}
    \centering
    \includegraphics{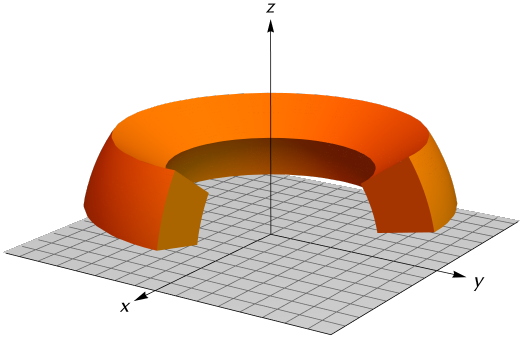}
    \caption{Volume of integration, with one quadrant cut away in order to show the cross section. In this illustration ${r=0.5}$, ${\beta=0.4}$, and ${d=0.1}$; normally, $d$ will be relatively small compared to $r$.}
    \label{fig:solid}
\end{figure}

The probability that the meteoroid is within a given volume is determined by integrating eq.\,\eqref{eq:fsx} over that volume. For a given heliocentric distance, ecliptic latitude, and characteristic distance ($d$), we choose to integrate over the toroid bounded by ${[r-d, \, r+d]}$ and ${[\beta - \frac{d}{r}, \, \beta + \frac{d}{r}]}$ (see Fig.\,\ref{fig:solid}). When $d$ is small compared to $r$, the volume of this region is:
\begin{align}
    V(r, \beta, d) &\simeq \begin{cases}
        8 \pi d^2 r \cos \beta & 
            \lvert \beta \rvert \le \frac{\pi}{2} - \frac{d}{r}  \\
        2 \pi d r^2 \left( \frac{\pi}{2} - \lvert \beta \rvert - \frac{d}{r} \right)^2 & 
            \lvert \beta \rvert > \frac{\pi}{2} - \frac{d}{r}
    \end{cases}
\end{align}
The second case in the above equation applies when the target location is close to one of the ecliptic poles and the volume of integration is shaped more like a cap than a toroid. Appendix\,\ref{sec:vol} provides equations for calculating both approximate and exact values of ${V(r, \beta, d)}$ in both cases. Note that ${V(r, \beta, d)}$ is finite for all locations with ${r > 0}$.

The choice of $d$ is somewhat arbitrary. We choose ${d = 0.02}$ for use in MEM~3.1: this is equivalent to 0.02~au at ${r = 1}$~au. This choice is motivated by our assumption that the sporadic meteoroid flux does not vary significantly within the Earth's Hill sphere (aside from the effects of the Earth and Moon). In this section, however, we will use ${d = 0.1}$ to generate some of the graphics in this section, so that the role of the characteristic distance can be more easily visualized. 

The probability that the meteoroid lies within the given volume at any point in time is 
\begin{align}
    P(r, \beta, d) &= 
        \left( F_s(s_2) - F_s(s_1) \right) \times
        \left( F_\xi(\xi_2) - F_\xi(\xi_1) \right) \, ,
\end{align}
where we use $F$ to denote the cumulative distribution function (CDF) of each variable. The quantities $s_1$ and $s_2$ correspond to our bounds on heliocentric distance (${r-d}$ and ${r+d}$, respectively) and $\xi_1$ and $\xi_2$ correspond to our bounds on ecliptic latitude (${\beta \pm d/r}$).

\begin{figure}
    \centering
    \includegraphics{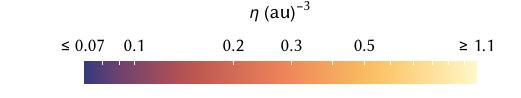}
    \includegraphics{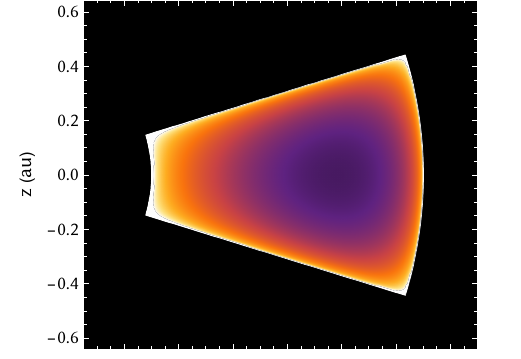}
    \includegraphics{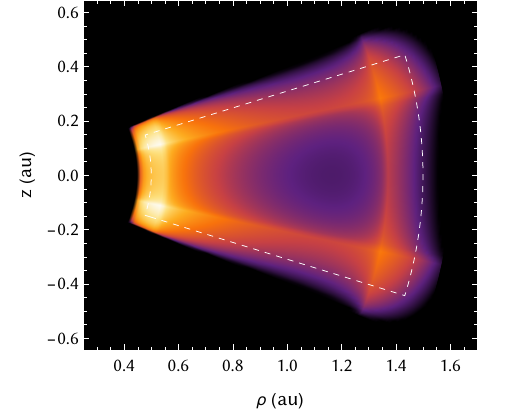}

    \vspace{6pt}
    
    \caption{Number density ($\eta$) of a single particle on an orbit with ${a=1}$, ${e=0.5}$, and ${i=0.3}$~radians. Heliocentric distance has been decomposed into an ecliptic component (${\rho = r \cos \beta}$) and an out-of-ecliptic component (${z = r \sin \beta}$). In the top panel, we use eq.\,\eqref{eq:limd0} to calculate $\eta$; it is not possible to show the full range of values in this case, as eq.\,\eqref{eq:limd0} is unbound near the edges of the depicted region. In the bottom panel, we use eq.\,\eqref{eq:nrbd}, selecting a fairly large value of ${d = 0.1 r}$ in order to better illustrate the differences between the two approaches. The dashed line in the lower panel shows the limits on the orbiting particle's position.}
    \label{fig:maps}
\end{figure}

Our dimensionless variables have CDFs with fairly simple closed forms:
\begin{align}
    F_s(s) &= \frac{1}{2} + \frac{1}{\pi} \left( 
        \sin^{-1} s - e \sqrt{1 - s^2} \right) \\
    F_\xi(\xi) &= \frac{1}{2} + \frac{1}{\pi} \sin^{-1} \xi \, .
\end{align}
\citep[See also appendix~A of][]{kessler81}.
Using these CDFs, we can calculate the number density at the desired location:
\begin{align}
    \eta(r, \beta, d) &= \frac{P(r, \beta, d)}{V(r, \beta, d)} \, .
    \label{eq:nrbd}
\end{align}
This equation is equivalent to eq.~31 of \cite{steel85}.
The limiting behavior of eq.\,\eqref{eq:nrbd} as the volume shrinks to zero is
\begin{align}
    \eta_0(r, \beta) &= \lim_{d \rightarrow 0} \eta(r, \beta, d) = \frac{f_{s \xi}(s, \xi)}{2 \pi r} \, .
    \label{eq:limd0}
\end{align}
Figure\,\ref{fig:maps} compares this limit with the number density calculated over a fairly large volume.

\subsection{Boundary handling}

Equation~\eqref{eq:limd0} is frequently used to calculate the spatial density of meteoroids with randomized orbit angles $\omega$, $\Omega$, and $M$ \citep[e.g.,][]{jones04,wiegert09,pokorny22,pokorny24}; it can also used to calculate the probability of collision between an object on a circular orbit and one on any other bound orbit \citep[e.g.,][]{weissman07,sachse18,frantseva22}. If eq.\,\eqref{eq:limd0} is convolved with and integrated over a continuous distribution of orbital elements \citep[e.g.,][]{leinert83,divine93}, the resulting number density is usually finite. 

If eq.\,\eqref{eq:limd0} is instead applied to a collection of discrete meteoroid orbits, the number density is undefined when ${r = q}$, ${r = Q}$, or ${\beta = \pm i}$. One possible approach to handling this issue is to formally exclude ${r = q, Q}$ (equivalent to ${s = \pm 1}$) and ${\beta = \pm i}$ (equivalent to ${\xi = \pm 1}$) from the allowed range of values. We find this approach inadequate, as it still produces unrealistically high values of the number density at locations that are near these singularities. This issue was present in older versions of MEM; we partially resolved the issue in MEM~3.0 by applying a smoothing term to the denominator of the radial PDF \citep{moorhead20}.

In MEM~3.1, we use eq.\,\eqref{eq:nrbd} to calculate number density. \cite{kessler81} and \cite{steel85} suggest using eq.\,\eqref{eq:nrbd} only for locations that lie near singularities, and eq.\,\eqref{eq:limd0} otherwise. We found it difficult to choose an appropriate cutoff value for ${\lvert s \rvert}$ that ensured a smooth transition from eq.\,\eqref{eq:limd0} to eq.\,\eqref{eq:nrbd} for all values of $e$, however, and found it simpler to use eq.\,\eqref{eq:nrbd} at all locations. We choose ${d = 0.01 r}$; this arbitrary choice is roughly the diameter of the Earth's Hill sphere at 1~au.

We would like to make a final note that while eq.\,\eqref{eq:nrbd} produces number densities that are finite and continuous, Fig.\,\ref{fig:maps} shows that the \emph{gradient} of eq.\,\eqref{eq:nrbd} is not continuous. A smoother alternative to eq.\,\eqref{eq:nrbd} can be obtained by convolving eq.\,\eqref{eq:fsx} with a kernel. For instance, we found that the convolution of quadratic kernels in $s$ and $\xi$ with their PDFs yielded closed-form integrals. However, these closed form integrals involved a much larger number of square roots and arcsine terms, resulting in a more numerically intensive calculation.

In appendix~\ref{sec:equiv}, we show that our spatial probability distribution (eq.\,\ref{eq:fsx}) is equivalent to the probability of collision between two orbiting objects, one of which has zero eccentricity.

\subsection{Encounter geometry}

\begin{figure}
    \centering
    \includegraphics{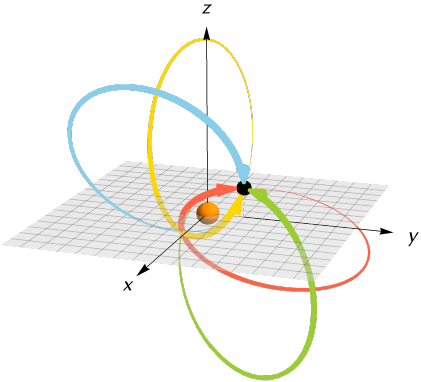}
    \vspace{0.1in}

    \caption{The four possible encounter geometries between a target (black circle) and a meteoroid with orbital elements ${a=3}$~au, ${e=0.8}$, and ${i = 60^\circ}$. 
    The target is 1.28~au from the Sun and is separated from the ecliptic plane by 0.8~au. The arrowheads indicate the meteoroid's direction of motion as it intercepts the target.}
    \label{fig:intersect}
\end{figure}

There are four sets of orbit angles that allow the meteoroid to intersect the target (see Fig.\,\ref{fig:intersect} for an illustration). These four sets of angles are those that satisfy
\begin{align}
    \cos \nu &= - \frac{e + s}{1 + e s} \, , ~ \, ~
    \sin (\omega + \nu) = \xi \, ,
    \label{eq:omnu}
\end{align}
where $\nu$ is true anomaly and $\omega$ is the argument of perihelion, measured in a Sun-centered ecliptic reference frame. In all four cases, the longitude of ascending node is uniquely determined by $\nu$ and $\omega$:
\begin{align}
    \sin \Omega &= - \frac{\sin (\omega + \nu)}{\cos \beta} \cos i \, , ~ \, ~
    \cos \Omega = \frac{\cos (\omega + \nu)}{\cos \beta} \, .
\end{align}
A derivation of the above equations, as well as an explicit enumeration of the four solutions of eq.\,\eqref{eq:omnu}, is presented in appendix\,\ref{sec:derivvel}. 

\cite{divine93} provides a solution for the velocity vector of the meteoroid at the target's location in both spherical and Cartesian coordinates (see Appendix~A of that work). However, we find it more convenient to express the velocity in cylindrical coordinates:
\begin{align}
    v_\rho &= - u \, e_s \cos \Omega - u \, e_c \sin \Omega \cos i \label{eq:vxo} \\
    v_\theta &= - u \, e_s \sin \Omega + u \, e_c \cos \Omega \cos i \label{eq:vyo} \\
    v_z &= u \, e_c \sin i \, . \label{eq:vzo} 
\end{align}
where the helper functions $u$, $e_s$, and $e_c$ are
\begin{align}
    u &= \sqrt{G M_\odot (1 - \beta_\text{rad})/(a(1-e^2))} \label{eq:u0} \\
    e_s &= e \sin \omega + \sin (\omega + \nu) \label{eq:es} \\
    e_c &= e \cos \omega + \cos (\omega + \nu) \, . \label{eq:ec} 
\end{align}
We use $\beta_\text{rad}$ to denote the ratio of radiation pressure to gravity \citep{burns79}. MEM assumes that ${\beta_\text{rad} \simeq 0}$ for the particles it models, which range in mass from 1~$\upmu$g to 10~g, or roughly 60~$\upmu$m to 1~cm in radius. However, $\beta_\text{rad}$ increases with decreasing particle size: it is roughly 1\% for 10~$\upmu$m particles, and can exceed 100\% for sub-micron particles \citep[which are then ``blown'' out of the solar system by radiation pressure;][]{burns79,moorhead21}. Our equations can be also be used when ${\beta_\text{rad} > 0}$; in this case, the reader must also ensure that this is taken into account when calculating the particle's semi-major axis and eccentricity.

Once the velocity vector ${v_\rho, v_\theta, v_z}$ has been calculated, it can be rotated about the $z$-axis to obtain $v_x$, $v_y$, and $v_z$:
\begin{align}
    v_x &= v_\rho \cos \lambda - v_\theta \sin \lambda \\
    v_y &= v_\rho \sin \lambda + v_\theta \cos \lambda \\
    v_z &= v_z \, , 
    \label{eq:vvec}
\end{align}
where $\lambda$ is the ecliptic longitude of the target's location:
\begin{align}
    \lambda &= \text{arctan2}(y, x) \, .
\end{align}

\subsection{Particle flux}

In this section, we outline MEM's approach to calculating the meteoroid flux from the spatial density and encounter velocity. These portions of MEM's algorithm have not been changed in version 3.1, but we describe them here for completeness.

Once the meteoroid's spatial density and velocity vector have been calculated using eqs.\,\ref{eq:nrbd} and \ref{eq:vvec}, we can calculate the corresponding flux ($\Phi$) relative to a moving target with heliocentric distance $r$ and ecliptic latitude $\beta$:
\begin{align}
    \Phi &= w \cdot 
    \eta(r, \beta, d) \cdot 
    \lVert \vec{v} - \vec{v}_\text{t} \rVert
    \label{eq:flux}
\end{align}
where $w$ is a weighting term, $\vec{v}$ is the velocity vector of the meteoroid given by eq.\,\ref{eq:vvec}, and $\vec{v}_\text{t}$ is the velocity vector of the target.
The weighting term is chosen so that when eq.\,\ref{eq:flux} is calculated using the Earth as target and summed over all modeled orbits, the result resembles the flux measured by in-situ experiments \citep{grun85} and by the Canadian Meteor Orbit Radar \citep{jones04,moorhead20}.

As the position and/or velocity of the target changes, the flux will also change. MEM integrates along the target's trajectory in a very simple manner: the sum of the particle flux contributed by all modeled meteoroid orbits is calculated for each target state vector input by the user. 

If the target location lies near a planet or Moon, the meteoroid's path may be bent by that body's gravity or even blocked from reaching the target location. Similarly, the flux may be enhanced by gravitational focusing or reduced by planetary shielding. These effects are handled using the algorithm outlined in \cite{staubach97}; see \cite{moorhead20} for additional discussion.

MEM bins the flux by impact direction and speed in order to facilitate the calculation of damage rates \citep{mem3tm}. 
The user can choose to receive the estimated flux relative to each target state vector, or they can opt to receive only the average. 
This approach offers a great deal of flexibility; no assumptions are made about the shape of the target's trajectory, the size or orientation of the target, or the target's vulnerability to impacts. On the other hand, the accuracy of the results will depend on the resolution of the user's input trajectory; see \cite{moorhead23} for guidance.

\section{Minor algorithmic improvements}

In addition to implementing the number density and impact geometry calculations described in section\,\ref{sec:math}, MEM~3.1 also repairs a bug in the calculation of average velocity. Finally, we updated MEM to handle each orbital population in parallel.

\subsection{Expanded heliocentric distance limit}
\label{sec:rhmax}

Historically, we have stated that MEM is valid between 0.2 and 2~au from the Sun \citep{mem2tm,mem3tm}. These limits are much more conservative than those of other meteoroid environment models: \cite{divine93}, for instance, quoted limits of 0.1 and 20~au for his model. This would appear to be at odds with the fact that the \cite{divine93} model is purely empirical (except for the asteroidal population, which is hypothetical), while MEM is a physics-based model that derives meteoroid orbital properties from parent body populations \citep{jones04}.

A recent internal review revealed that these distance limits may simply have been the minimum range required of the model, as is implied in \cite{mcnamara04}. In contrast, \cite{jones04} felt that the dynamical model underpinning MEM could potentially be used past Jupiter. 

No substantial changes are needed to expand MEM's heliocentric distance limits; MEM~3.0 simply checks the heliocentric distance of the user's trajectory and raises an error if it passes outside of MEM's bounds. In MEM~3.1, we have edited the outer limit to be 4.6~au rather than 2~au. This distance is obtained by subtracting Jupiter's Hill radius (0.34~au) from its perihelion distance (4.95~au) and rounding down to the nearest 0.1~au. 

We stop just short of Jupiter's sphere of influence for three reasons. The first reason is that MEM uses the \cite{staubach97} method of calculating the effects of planetary gravity on meteoroid flux and direction. This method assumes that meteoroids originating from a particular orbit intercept all parts of a planet's Hill sphere with the same velocity vector. This is approximately true for a planet like the Earth, but Jupiter's Hill sphere spans $8^\circ$ in ecliptic latitude. Further investigation is needed to determine the degree of inaccuracy introduced by applying the \cite{staubach97} method in such a case.

Our second reason for excluding Jupiter's orbit is that the giant planet is encircled by dust rings. These rings produced a three-orders-of-magnitude increase in the micron-sized particle flux measured by the Pioneer~10 and 11 spacecraft \citep{kruger04}, both of which passed within one Jovian radius of the planet's ``surface.'' Although most spacecraft, even those performing gravitational slingshots, do not pass through the four main dust rings, we will need to determine the maximum distance at which the rings can noticeably enhance the flux before including locations near Jupiter.

Finally, we do not extend MEM's limits \emph{past} Jupiter because the model does not include particles originating from Centaurs and Kuiper belt objects. The \cite{poppe19} model predicts that material from Centaurs and KBOs dominates the dust environment outside Jupiter's orbit. Thus, the expansion of MEM to the outer Solar System would require revision or replacement of the \cite{jones04} model.

At this point, we would also like to note that while the \cite{jones04} model includes asteroidal meteoroids, MEM~3 does not. Asteroidal meteoroids have lower eccentricities than cometary meteoroids, and these eccentricities will be further reduced by Poynting-Robertson drag \citep{burns79}. The Jones model predicts that by the time these particles encounter the Earth, their orbital eccentricities and geocentric velocities are so low that even a tiny flux is massively enhanced by gravitational focusing \citep{moorhead20}, yet does not match any observed dynamical population of meteors. For this reason, we excluded asteroidal meteoroids from MEM~3.

A number of models do include asteroidal meteoroids: the asteroidal component is usually assumed to contribute no more than 10\% of the mass flux onto the Earth \citep{nesvorny11,carrillo16,pokorny24}. The asteroidal population may be more significant at larger heliocentric distances, however. For instance, the Interplanetary Meteoroid Environment Model \citep[IMEM2][]{soja19} predicts that the number density of 250~$\upmu$m asteroidal particles is equal to that produced by JFCs at locations within the main asteroid belt. On the other hand, these particles are on low-eccentricity orbits compared to cometary meteoroids, and their encounter velocities with spacecraft will therefore also be low. This in turn means that the flux of asteroidal particles will be lower than that of cometary particles, and that the rate of damage, which depends on the kinetic energy of the impactor, will be lower still.

\subsection{Improved average speed calculation}
\label{sec:vavg}

In the course of updating the code with the changes outlined in section\,\ref{sec:math}, we uncovered and repaired a bug in MEM's average velocity calculations. 
MEM~3 uses the \cite{west79} algorithm to compute single-pass averages and standard deviations of meteoroid speed. Incorporating this algorithm into MEM~3 resulted in a version of the code that ran several times faster than the previous version and does not require any intermediate file storage \citep{moorhead20}. However, if the flux in a given bin is zero for the first state vector processed, the initial average speed is indeterminate (i.e., it is $\tfrac{0}{0}$) and \texttt{NaN} values appear in the run outputs. 

We have addressed this behavior by assigning an average speed of 0, with a corresponding weight of 0, to empty bins. This allows the code to correctly compute the average speed in bins that are initially empty.

\subsection{Parallelization}
\label{sec:parallel}

We added some simple parallelization to MEM~3.1 using the C++ \texttt{thread} class; each source population can be handled by a separate processor, so the total reduction in run time is between a factor of two and three. 
For instance, the \emph{Solaris} trajectory presented in \S\ref{sec:solaris} can be processed by MEM~3.1 in 3~minutes and 48~seconds, while the same trajectory takes 8~minutes and 14~seconds to process with MEM~3.0. These times were obtained by running both versions of MEM in high-fidelity mode, and the trajectory file contained 338 state vectors.

\section{Results}
\label{sec:results}

\subsection{Comparison with MEM~3.0}
\label{sec:mem30}

When a spacecraft or observer lies near the ecliptic plane, the assumptions of \cite{jones04} are valid and therefore MEM~3.1 and MEM~3.0 should yield similar results. To test this, we ran the trajectory ``ISSExample.txt'' through both versions of the code. This trajectory contains ten state vectors for the International Space Station (ISS) and is part of the MEM~3.0 desktop package.\footnote{https://software.nasa.gov/software/MFS-32205-2} This file is not useful for risk assessments -- ten state vectors are certainly not enough to fully sample the spacecraft's orbit -- but is useful as an example of the trajectory file format that MEM requires. It is also a useful test case here: because the ISS orbits the Earth, it necessarily remains close to the ecliptic plane. 

\begin{figure}
    \centering
    \includegraphics[width=\linewidth]{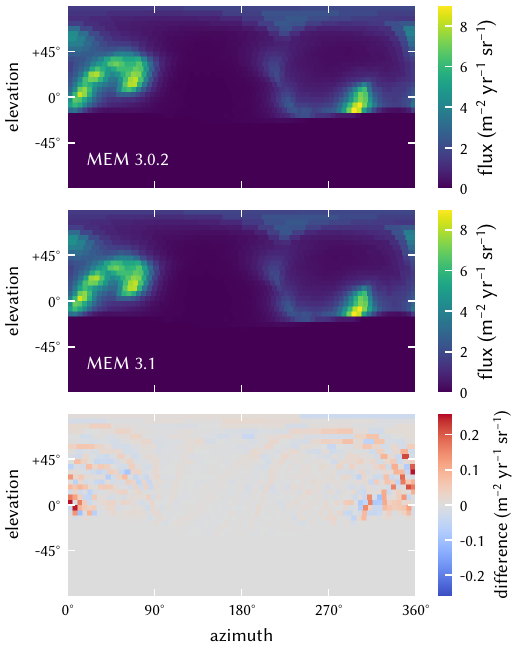}
    \caption{The flux of meteoroids larger than 1~$\upmu$g relative to the ISS, averaged over 10 sample state vectors using MEM~3.0 and MEM~3.1. The flux is binned by azimuth and elevation angle in the velocity-normal-binormal (VNB) reference frame, and therefore the flux is quoted per steradian. The bottom panel shows the difference between the two sets of results.}
    \label{fig:diff}
\end{figure}

The two runs do not produce identical results, because MEM~3.0 uses the equivalent of eq.\,\eqref{eq:limd0} to calculate number density, while MEM~3.1 uses eq.\,\eqref{eq:nrbd} to average the number density over a distance of 0.02~r (which is equal to 0.02~au in this case). The differences are quite small, however; in Fig.\,\ref{fig:diff}, we see that the largest difference in flux per angular bin is less than 3\% of the flux scale. Larger choices of $d$ would produce more smoothing of the middle panel of Fig.\,\ref{fig:diff}.

We have performed a few additional runs, and find that MEM~3.1 generates flux values that are approximately 1\% higher on average than those computed using MEM~3.0. This is well within the uncertainty in the meteoroid environment, which is typically estimated to be a factor between two and three \citep{grun13,moorhead20,moorhead24}.

\subsection{Comparison with zodiacal light}
\label{sec:zody}

The variation in intensity of zodiacal light with ecliptic latitude provides a constraint on the inclination distribution of dust particles \citep{nesvorny11}.

We will follow the example of \citep{nesvorny06} and \citep{nesvorny10} in simulating thermal emission from our modeled meteoroids and comparing the result with IRAS observations. The flux emitted by a meteoroid and intercepted by an observer is:
\begin{align}
F(\lambda) &= \epsilon(\lambda, p) \, B(\lambda, T) \frac{p^2}{\ell^2} \, , \text{where} \\
B(\lambda, T) &= \frac{2 \pi h c^2}{\lambda^5} \left( e^{h c/\lambda k T} - 1 \right)^{-1}
\end{align}
where $\epsilon$ is the emissivity, $p$ is particle radius, $\ell$ is geocentric distance, $\lambda$ is wavelength, and $T$ is temperature. The remaining quantities are the Planck constant ($h$), the speed of light ($c$), and the Boltzmann constant ($k$). The energy flux, $B$, depends on the equilibrium temperature, $T$. We follow \cite{nesvorny10} in adopting
\begin{align}
    T &\simeq 280~\text{K} \, \left(\frac{r}{\text{1 au}}\right)^{-1/2} \, .
\end{align}

MEM assumes that the particle size distribution and orbital distribution are independent within the modeled mass range (1~$\upmu$g--10~g).
We therefore focus on reproducing the \emph{shape} of the IRAS profile, rather than its magnitude, and ignore size-dependent terms such as the emissivity. Our only consideration of particle size is to select the IRAS profile that is most relevant to larger particles (25~$\upmu$m wavelength) for comparison with MEM. For a given size, the relative apparent brightness of a particle is:
\begin{align}
    b(r, \ell) &= \frac{1}{ e^{\sqrt{r/r_0}} - 1 }
    \cdot
    \frac{1}{\ell^2}
\end{align}
where ${r_0 = (h c / \lambda k (\text{280 K}))^2~\text{au} = 4.23~\text{au}}$ for $\lambda = 25~\upmu$m.

The profiles presented in \cite{nesvorny10} correspond to a solar elongation angle of 90$^\circ$ when viewed from IRAS, which orbits the Earth. This corresponds to those particles whose Earth-particle vector is perpendicular to the Sun-Earth vector. Thus,
\begin{align}
    r^2 &= \ell^2 + (\text{1 au})^2 \, \text{ and } \\
    r \sin \beta &= \ell \sin \phi \, ,
\end{align}
where $\phi$ is geocentric ecliptic latitude. For a given value of $\phi$, the volume of integration is ${\ell^2 \, d \ell}$, and therefore the expected relative brightness is
\begin{align}
    \int_{\ell_\text{min}}^{\ell_\text{max}}
    \eta(r(\ell), \beta(\ell, \phi)) \cdot
    b(r(\ell), \ell) \cdot
    \ell^2 \, d \ell \, ,
    \label{eq:bright}
\end{align}
where $\ell_\text{min}$ and $\ell_\text{max}$ are the minimum and maximum possible geocentric distance of the particle. Equation~\eqref{eq:bright} is comparable to eq.\,(6) of \cite{nesvorny10}. However, we sum over the orbits in our model rather than integrate over a distribution of orbital elements, and we drop size-dependent terms as discussed above. We use eq.\,\eqref{eq:limd0} to calculate $\eta$, and evaluate eq.\,\eqref{eq:bright} using SciPy's implementation of the QUADPACK numerical integration library \citep{scipy,piessens83}.

\begin{figure}
    \centering
    \includegraphics[width=\linewidth]{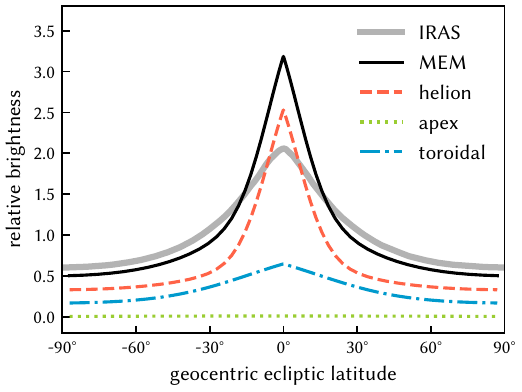}
    \includegraphics[width=\linewidth]{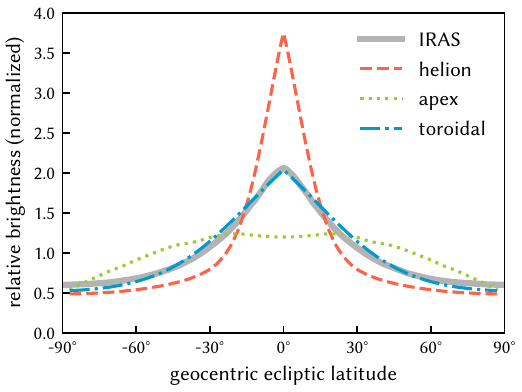}
    \caption{The vertical profile of zodiacal brightness observed by IRAS at ${25 \upmu}$m wavelength and a solar elongation angle of $90^\circ$ and that corresponding to MEM's meteoroid orbits. In addition to the overall brightness profile, we also show the contribution from each of MEM's populations (top); in the lower panel, we normalize each population separately. The IRAS profile was extracted from Fig.~2 of \cite{nesvorny10}, mirrored across $0^\circ$, and averaged. The units of relative brightness are arbitrary.}
    \label{fig:emit}
\end{figure}

Figure\,\ref{fig:emit} compares MEM's predicted particle brightness profile with the 25-micron IRAS profile from \cite{nesvorny10}. We see that the overall profile is more sharply peaked at the ecliptic than the IRAS data. 

\subsection{Application to \emph{Solaris}}
\label{sec:solaris}

\begin{figure}
    \centering
    \includegraphics[width=\linewidth]{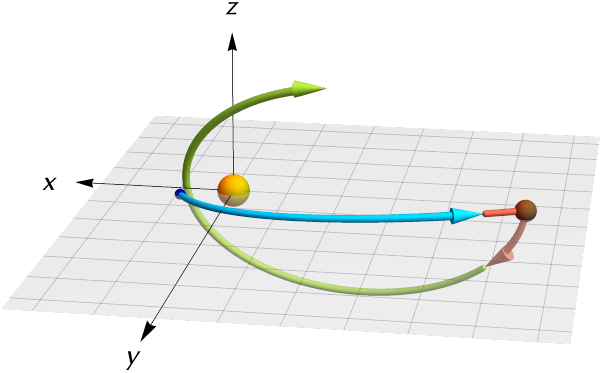}
    \caption{Solaris's trajectory, showing its initial transfer from Earth to Jupiter (blue), gravitational slingshot (red), and subsequent high-inclination orbit (green). The grid lines lie in the ecliptic plane and are spaced at 1~au intervals. The red portion of the trajectory indicates when the spacecraft is more than 4.6~au from the Sun and thus is out of MEM~3.1's range.}
    \label{fig:solaris}
\end{figure}

In this section, we use part of \emph{Solaris}'s trajectory to illustrate the significance of incorporating these algorithms into MEM. \emph{Solaris} initially remains close to the ecliptic after its launch. It then uses a Jupiter gravity assist to transition into a high-inclination (${i \sim 75^\circ}$) heliocentric orbit; Fig.\,\ref{fig:solaris} contains a diagram of this trajectory.

We restrict our comparison to the portion of the trajectory in which the spacecraft has already completed its Jupiter gravity assist, it lies within 2~au of the Sun, and its ecliptic latitude is monotonically increasing. This ensures that the modeled trajectory lies within the heliocentric distance limits of MEM~3.0 and that the resulting plot is relatively simple. 

\begin{figure}
    \centering
    \includegraphics[width=\linewidth]{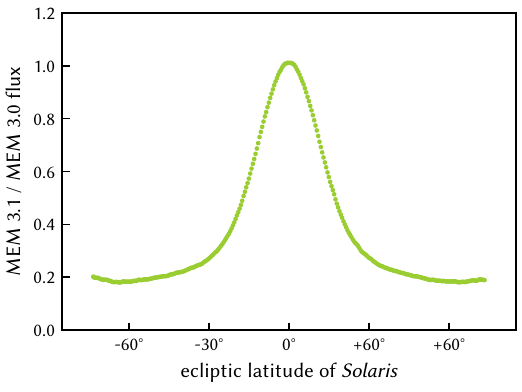}
    \caption{The ratio of the meteoroid flux onto Solaris predicted by MEM~3.1 to that predicted by MEM~3.0. We have sampled the trajectory between the point of its minimum ecliptic latitude to its maximum ecliptic latitude. The points shown are not evenly spaced in time.}
    \label{fig:flux}
\end{figure}

Figure\,\ref{fig:flux} compares the meteoroid flux predicted by MEM~3.0 and 3.1. We see again that the two models predict similar flux values when the spacecraft is near the ecliptic. However, when Solaris is at ecliptic latitudes between ${\sim 30^\circ}$ and $60^\circ$, the flux is about one-fifth what it would be if the spacecraft were moving at a similar speed near the ecliptic. At these latitudes, the spacecraft has moved out of range of the helion/antihelion particles. 

\section{Possible generalizations}

It is straightforward (if not necessarily easy) to modify our derivation of number density to allow non-uniform probability distributions for $\omega$ and $\Omega$.
We discuss these cases as well as a few additional possible generalizations in this section.

\paragraph{Non-random $\bm{\Omega}$.} 
For short-period meteoroid streams, the longitude of ascending node, $\Omega$, may be the last orbit angle to be randomized \citep{babadzhanov08a,babadzhanov08b,asher93}. 
If $\omega$ and $M$ are randomized \emph{and} are uncorrelated with $\Omega$, then generalizing our approach to a non-uniform distribution of $\Omega$ is fairly simple, as neither $f_s$ nor $f_\xi$ depend on $\Omega$. 

The probability distribution of $\Omega$ will, however, affect the likelihood of the four different encounter probabilities. This can be accounted for by weighting the nominal spatial density by ${
f_\Omega(\Omega_i)}$ for each value of $\Omega_i$ that intersects the target's position:
\begin{align}
    \eta_\Omega(r, \beta, d) &\propto 
    \eta(r, \beta, d) \times
    \sum_{j=1}^4 f_\Omega(\Omega_j(s, \xi))
\end{align}
We suggest that this approach could potentially be used to build an empirical model of the seasonal variations in sporadic flux reported by \cite{cb06}.

\paragraph{Non-random $\bm{\omega}$.} 
If the argument of pericenter, $\omega$, is not randomly or evenly distributed between 0 and ${2 \pi}$, then eq.\,\ref{eq:fnw} does not follow from eq.\,\ref{eq:fnu}. The form of $\eta$ will depend on the form of $f_\omega(\omega)$.

\paragraph{Non-constant $\bm{a}$, $\bm{e}$, or $\bm{i}$.} The primary orbital elements of a meteoroid's orbit change over time with nearly all forms of orbital evolution, including planetary perburbations, Poynting-Robertson drag, and Kozai-Lidov cycles \citep[see][for a treatment of the latter]{vokrouhlicky12,pokorny13}. So long as these elements are uncorrelated with $\omega$ or $M$ on a population level, our general approach may still be used; the user need only weight $\eta$ or $\eta_\Omega$ by the PDF of $a$, $e$, and/or $i$. Thus, our approach can be used on a collection of sporadic meteoroids that are subject to Poynting-Robertson drag (which affects $a$, $e$, and $\omega$), but whose values of $\omega$ are expected to be randomized and uncorrelated with $a$ and $e$.

\section{Conclusions}

This paper describes the development of a new version of NASA's Meteoroid Engineering Model (MEM) -- version 3.1 -- that is capable of describing the meteoroid flux at high ecliptic latitudes. We present an algorithm for calculating meteoroid spatial density from a representative modeled orbit that is equivalent to that of \cite{kessler81}, but has been volume-averaged to produce a result that is finite at all locations and varies smoothly between locations. We calculate the meteoroid's speed and direction in cylindrical coordinates, which we find more convenient for our rotationally symmetric sporadic model. These algorithms have also been implemented in \texttt{Python} and made publicly available.

Using MEM~3.1, we find that the meteoroid flux drops significantly with ecliptic latitude. The proposed \emph{Solaris} solar observation mission, for instance, experiences an approximate 80\% reduction in meteoroid flux at its highest ecliptic latitudes compared to the near-ecliptic flux. 

In addition to expanding MEM's range in terms of ecliptic latitude, we have also expanded the maximum heliocentric distance from 2~au to 4.6~au. Although this greatly expands the volume of space covered by MEM, the model's use remains restricted to locations that lie interior to Jupiter. Additional modeling work is required to expand MEM to include locations near Jupiter (including an improved treatment of gravitational focusing and the possible inclusion of particles in Jupiter's rings) and beyond \citep[such as the inclusion of additional meteoroid-producing populations;][]{poppe19,pokorny22}.

We also compare MEM's orbital populations to the vertical zodiacal light profile from IRAS presented in \cite{nesvorny10}. We find that MEM's meteoroids appear to be more concentrated near the ecliptic than is consistent with the IRAS data. These data provide a useful constraint on the inclination distribution of meteoroids and should be incorporated into any future efforts to improve MEM's orbital populations.

\vspace{\baselineskip}

The authors would like to thank Noble Hatten for providing the \emph{Solaris} trajectory used in this paper. PP acknowledges support provided by NASA’s Planetary Science Division Research Program, through ISFM work packages EIMM and Planetary Geodesy at NASA Goddard Space Flight Center, NASA award numbers 80GSFC24M0006 and 80NSSC21K0153.

\bibliography{refs}

@software{repo,
  author       = {Moorhead, Althea},
  title        = {metnumdens},
  year         = 2025,
  publisher    = {Zenodo},
  version      = {1.0},
  doi          = {10.5281/zenodo.17610831},
  url          = {https://github.com/nasa/metnumdens},
}

@article{asher93,
    author = {{Asher}, D.~J. and {Clube}, S.~V.~M.},
    title = "{An extraterrestrial influence during
        the current glacial-interglacial}",
    journal = {Quarterly Journal of the RAS},
    year = 1993,
    volume = {34},
    pages = {481-511},
}

@article{babadzhanov08a,
    author = {{Babadzhanov}, P.~B. and {Williams},
        I.~P. and {Kokhirova}, G.~I.},
    title = "{Meteor showers associated with
        2003EH1}",
    journal = {MNRAS},
    year = 2008,
    volume = {386},
    number = {4},
    pages = {2271-2277},
    doi = {10.1111/j.1365-2966.2008.13202.x},
}

@article{babadzhanov08b,
    author = {{Babadzhanov}, P.~B. and {Williams},
        I.~P. and {Kokhirova}, G.~I.},
    title = "{near-earth asteroids among the piscids
        meteoroid stream}",
    journal = {Astronomy and Astrophysics},
    year = 2008,
    volume = {479},
    number = {1},
    pages = {249-255},
    doi = {10.1051/0004-6361:20078185},
}

@article{bottke20,
    author = {{Bottke}, W.~F. and {Moorhead}, A.~V.
        and {Connolly}, H.~C. and
        {Hergenrother}, C.~W. and {Molaro},
        J.~L. and {Michel}, P. and {Nolan},
        M.~C. and {Schwartz}, S.~R. and
        {Vokrouhlick{\'y}}, D. and {Walsh},
        K.~J. and {Lauretta}, D.~S.},
    year = 2020,
    title = "{Meteoroid impacts as a source of
        Bennu's particle ejection events}",
    journal = {Journal of Geophysical Research
        (Planets)},
    volume = {125},
    number = {8},
    eid = {e06282},
    pages = {e06282},
    doi = {10.1029/2019JE006282},
}

@article{burns79,
    author = {{Burns}, J.~A. and {Lamy}, P.~L. and
        {Soter}, S.},
    title = "{Radiation forces on small particles in
        the solar system}",
    journal = {Icarus},
    year = 1979,
    volume = {40},
    number = {1},
    pages = {1-48},
    doi = {10.1016/0019-1035(79)90050-2},
}

@article{carrillo16,
    author = {{Carrillo-S{\'a}nchez}, J.~D. and
        {Nesvorn{\'y}}, D. and {Pokorn{\'y}}, P.
        and {Janches}, D. and {Plane}, J.~M.~C.},
    title = "{Sources of cosmic dust in the Earth's
        atmosphere}",
    journal = {Geophysics Research Letters},
    year = 2016,
    volume = {43},
    number = {23},
    pages = {11,979-11,986},
    doi = {10.1002/2016GL071697},
}

@article{cb06,
    author = {{Campbell-Brown}, M.~D. and {Jones},
        J.},
    title = "{Annual variation of sporadic radar
        meteor rates}",
    journal = {MNRAS},
    year = 2006,
    volume = {367},
    number = {2},
    pages = {709-716},
    doi = {10.1111/j.1365-2966.2005.09974.x},
}

@article{divine93,
    author = {{Divine}, Neil},
    title = "{Five populations of interplanetary
        meteoroids}",
    journal = {Journal of Geophysics Research},
    year = 1993,
    volume = {98},
    number = {E9},
    pages = {17029-17048},
    doi = {10.1029/93JE01203},
}

@article{frantseva22,
    author = {{Frantseva}, Kateryna and
        {Nesvorn{\'y}}, David and {Mueller},
        Michael and {van der Tak}, Floris F.~S.
        and {ten Kate}, Inge Loes and
        {Pokorn{\'y}}, Petr},
    title = "{Exogenous delivery of water to
        Mercury}",
    journal = {Icarus},
    year = 2022,
    volume = {383},
    eid = {114980},
    pages = {114980},
    doi = {10.1016/j.icarus.2022.114980},
}

@inproceedings{grun13,
    author = {{Gr\"{u}n}, Eberhard and {Srama}, Ralf and
        {Hor\'{a}nyi}, Mih\'{a}ly and {Kr\"{u}ger}, Harald
        and {Soja}, Rachel and {Sterken}, Veerle
        and {Sternovsky}, Zotan and {Strub},
        Peter},
    title = "{Comparative analysis of the ESA and
        NASA interplanetary meteoroid enviroment
        models}",
    booktitle = {6th European Conference on Space
        Debris},
    year = 2013,
    editor = {{Ouwehand}, L.},
    series = {ESA Special Publication},
    volume = {723},
    eid = {36},
    pages = {36},
}

@article{grun85,
    author = {{Gr\"{u}n}, E. and {Zook}, H.~A. and
        {Fechtig}, H. and {Giese}, R.~H.},
    title = "{Collisional balance of the meteoritic
        complex}",
    journal = {Icarus},
    year = 1985,
    volume = {62},
    number = {2},
    pages = {244-272},
    doi = {10.1016/0019-1035(85)90121-6},
}

@inproceedings{hassler23,
    author = {{Hassler}, Donald M. and {Gibson},
        Sarah E. and {Newmark}, Jeffrey S. and
        {Featherstone}, Nicholas Andrew and
        {Viall}, Nicholeen M. and {Upton}, Lisa
        A. and {Hoeksema}, J. Todd and
        {Auch{\`e}re}, Fr{\'e}d{\'e}ric and
        {Birch}, Aaron and {Braun}, Douglas C.
        and {Charbonneau}, Paul and {Colannino},
        Robin and {DeForest}, Craig and
        {Dikpati}, Mausumi and {Downs}, Cooper
        and {Duncan}, Nicole and {Elliott},
        Heather Alison and {Fan}, Yuhong and
        {Fineschi}, Silvano and {Gizon}, Laurent
        and {Gosain}, Sanjay and {Harra}, Louise
        and {Hindman}, Bradley and {Berghmans},
        David and {Lepri}, Susan T. and
        {Linker}, Jon and {Moldwin}, Mark B. and
        {Munoz-Jaramillo}, Andres and {Nandy},
        Dibyendu and {Rivera}, Yeimy and
        {Schou}, Jesper and {Sokol}, Justyna and
        {Thompson}, Barbara J. and {Velli},
        Marco and {Woods}, Thomas N. and {Zhao},
        Junwei},
    title = "{Solaris: A focused solar polar
        discovery-class mission to achieve the
        highest priority heliophysics science
        now}",
    booktitle = {Bulletin of the American
        Astronomical Society},
    year = 2023,
    volume = {55},
    eid = {164},
    pages = {164},
    doi = {10.3847/25c2cfeb.408d006f},
}

@book{hogg20,
    author = {Hogg, Robert V. and 
        Tanis, Elliot A. and Zimmerman, Dale L.},
    year = {2018},
    title = {Probability and statistical inference},
    publisher = {Pearson},
    address = {New York, NY},
    isbn = {9780135189399}
}

@techreport{jones04,
    author = {{Jones}, J.},
    title = "{Meteoroid Engineering Model -- Final Report}",
    institution = {NASA Space Environment Effects Program},
    number = {CR-2004-400},
    year = 2004,
    url = {https://ntrs.nasa.gov/citations/20190030409}
}

@article{kessler81,
    author = {{Kessler}, D.~J.},
    title = "{Derivation of the collision probability
        between orbiting objects: The lifetimes
        of Jupiter's outer moons}",
    journal = {Icarus},
    year = 1981,
    volume = {48},
    number = {1},
    pages = {39-48},
    doi = {10.1016/0019-1035(81)90151-2},
}

@incollection{kruger04,
    author = {{Kr{\"u}ger}, Harald and {Hor{\'a}nyi},
        Mih{\'a}ly and {Krivov}, Alexander V.
        and {Graps}, Amara L.},
    title = "{Jovian dust: Streams, clouds and
        rings}",
    booktitle = {Jupiter: The Planet, Satellites and
        Magnetosphere},
    year = 2006,
    editor = {{Bagenal}, Fran and {Dowling}, Timothy
        E. and {McKinnon}, William B.},
    publisher = {Cambridge University Press},
    volume = {1},
    pages = {219-240},
}

@article{leinert83,
    author = {{Leinert}, C. and {Roser}, S. and
        {Buitrago}, J.},
    title = "{How to maintain the spatial
        distribution of interplanetary dust}",
    journal = {Astronomy and Astrophysics},
    year = 1983,
    volume = {118},
    number = {2},
    pages = {345-357},
}

@article{mcnamara04,
    author = {{McNamara}, H. and {Jones}, J. and
        {Kauffman}, B. and {Suggs}, R. and
        {Cooke}, W. and {Smith}, S.},
    title = "{Meteoroid Engineering Model (MEM): A
        meteoroid model for the inner solar
        system}",
    journal = {Earth Moon and Planets},
    year = 2004,
    volume = {95},
    number = {1-4},
    pages = {123-139},
    doi = {10.1007/s11038-005-9044-8},
}

@article{memioc,
    author = {{Moorhead}, Althea V.},
    title = "{Meteoroid Engineering Model (MEM) 3: NASA's newest meteoroid model}",
    journal = {Proceedings of the First International Orbital Debris Conference, Houston},
    year = {2019},
    volume = {article no. 6054},
    pages = {1-8},
    url = {https://www.hou.usra.edu/meetings/orbitaldebris2019/orbital2019paper/pdf/6054.pdf},
}

@techreport{mem2tm,
    author = {{Moorhead}, Althea V. and {Koehler}, H.~M. and {Cooke}, W.~J.},
    title = "{NASA Meteoroid Engineering Model Release 2.0}",
    institution = {NASA},
    number = {TM-2015-218214},
    year = 2015,
    url = {https://ntrs.nasa.gov/citations/20150021449},
}

@techreport{mem3tm,
    author = {{Moorhead}, Althea V.},
    title = "{NASA Meteoroid Engineering Model (MEM) Version 3}",
    institution = {NASA},
    number = {TM-2020-220555},
    year = 2020,
    url = {https://ntrs.nasa.gov/citations/20200000563},
}

@article{moorhead20,
    author = {{Moorhead}, Althea V. and 
        {Kingery}, Aaron and {Ehlert}, Steven},
    year = 2020,
    title = "{NASA's Meteoroid Engineering Model 3 
        and its ability to replicate spacecraft 
        impact rates}",
    journal = {Journal of Spacecraft and Rockets},
    volume = {57},
    pages = {160-176},
    doi = {10.2514/1.A34561},
}

@article{moorhead21,
    author = {{Moorhead}, Althea V.},
    title = "{Forbidden mass ranges for shower
        meteoroids}",
    journal = {Icarus},
    year = 2021,
    volume = {354},
    eid = {113949},
    pages = {113949},
    doi = {10.1016/j.icarus.2020.113949},
}

@article{moorhead23,
    author = {{Moorhead}, Althea V. and {Milbrandt},
        Katie and {Kingery}, Aaron},
    title = "{A library of meteoroid environments
        encountered by spacecraft in the inner
        solar system}",
    journal = {Advances in Space Research},
    year = 2023,
    volume = {72},
    number = {10},
    pages = {4582-4595},
    doi = {10.1016/j.asr.2023.08.016},
}

@techreport{moorhead24,
    author = {{Moorhead}, A.~V. and {Campbell-Brown}, M.~D. and {Brown}, P.~G.},
    title = "{The Activity Profiles and Peak Flux of Radar Meteor Showers}",
    institution = {NASA Meteoroid Environment Office},
    number = {OSMA/MEO/Report–13},
    year = 2024,
    url = {https://ntrs.nasa.gov/citations/20240005599}
}

@article{muller20,
    author = {{M{\"u}ller}, D. and {St. Cyr}, O.~C.
        and {Zouganelis}, I. and {Gilbert},
        H.~R. and {Marsden}, R. and
        {Nieves-Chinchilla}, T. and {Antonucci},
        E. and {Auch{\`e}re}, F. and
        {Berghmans}, D. and {Horbury}, T.~S. and
        {Howard}, R.~A. and {Krucker}, S. and
        {Maksimovic}, M. and {Owen}, C.~J. and
        {Rochus}, P. and {Rodriguez-Pacheco}, J.
        and {Romoli}, M. and {Solanki}, S.~K.
        and {Bruno}, R. and {Carlsson}, M. and
        {Fludra}, A. and {Harra}, L. and
        {Hassler}, D.~M. and {Livi}, S. and
        {Louarn}, P. and {Peter}, H. and
        {Sch{\"u}hle}, U. and {Teriaca}, L. and
        {del Toro Iniesta}, J.~C. and
        {Wimmer-Schweingruber}, R.~F. and
        {Marsch}, E. and {Velli}, M. and {De
        Groof}, A. and {Walsh}, A. and
        {Williams}, D.},
    title = "{The Solar Orbiter mission - Science overview}",
    journal = {Astronomy and Astrophysics},
    year = 2020,
    volume = {642},
    eid = {A1},
    pages = {A1},
    doi = {10.1051/0004-6361/202038467},
}

@book{murray99,
    author = {{Murray}, Carl D. and {Dermott}, Stanley F.},
    year = 2000,
    title = "{Solar System Dynamics}",
    publisher = {Cambridge University Press},
    address = {New York, N.Y.},
    isbn = {978-0-521-57597-4},
}

@article{nesvorny06,
    author = {{Nesvorn{\'y}}, David and
        {Vokrouhlick{\'y}}, David and {Bottke},
        William F. and {Sykes}, Mark},
    year = 2006,
    title = "{Physical properties of asteroid dust
        bands and their sources}",
    journal = {Icarus},
    volume = {181},
    number = {1},
    pages = {107-144},
    doi = {10.1016/j.icarus.2005.10.022},
}

@article{nesvorny10,
    author = {{Nesvorn{\'y}}, David and {Jenniskens},
        Peter and {Levison}, Harold F. and
        {Bottke}, William F. and
        {Vokrouhlick{\'y}}, David and
        {Gounelle}, Matthieu},
    title = "{Cometary origin of the zodiacal cloud
        and carbonaceous micrometeorites.
        Implications for hot debris disks}",
    journal = {Astrophysical Journal},
    year = 2010,
    volume = {713},
    number = {2},
    pages = {816-836},
    doi = {10.1088/0004-637X/713/2/816},
}

@article{nesvorny11,
    author = {{Nesvorn{\'y}}, David and {Janches},
        Diego and {Vokrouhlick{\'y}}, David and
        {Pokorn{\'y}}, Petr and {Bottke},
        William F. and {Jenniskens}, Peter},
    year = 2011,
    title = "{Dynamical model for the zodiacal cloud
        and sporadic meteors}",
    journal = {Astrophysical Journal},
    volume = {743},
    number = {2},
    eid = {129},
    pages = {129},
    doi = {10.1088/0004-637X/743/2/129},
}

@article{opik51,
    author = {{Opik}, E.~J.},
    year = 1951,
    title = "{Collision probabilities with the planets
        and the distribution of interplanetary
        matter}",
    journal = {Proceedings of the Royal Irish Academy. Section A: Mathematical and Physical Sciences},
    volume = {54},
    pages = {165-199},
}

@book{piessens83,
    author = {{Piessens}, Robert and 
        {de Doncker-Kapenga}, Elise and {\"{U}berhuber}, Christoph W. and {Kahaner}, David},
    year = 1983,
    title = "{QUADPACK: A subroutine package for automatic integration}",
    publisher = {Springer-Verlag},
    address = {Heidelberg},
    doi = {10.1007/978-3-642-61786-7}
}

@article{pokorny13,
    author = {{Pokorn{\'y}}, Petr and
        {Vokrouhlick{\'y}}, David},
    title = "{{\"O}pik-type collision probability for
        high-inclination orbits: Targets on
        eccentric orbits}",
    journal = {Icarus},
    year = 2013,
    volume = {226},
    number = {1},
    pages = {682-693},
    doi = {10.1016/j.icarus.2013.06.015},
}

@article{pokorny22,
    author = {{Pokorn{\'y}}, Petr and {Szalay}, Jamey
        R. and {Hor{\'a}nyi}, Mih{\'a}ly and
        {Kuchner}, Marc J.},
    title = "{Modeling meteoroid impacts on the Juno
        spacecraft}",
    journal = {Planetary Science Journal},
    year = 2022,
    volume = {3},
    number = {1},
    eid = {14},
    pages = {14},
    doi = {10.3847/PSJ/ac4019},
}

@article{pokorny24,
    author = {{Pokorn{\'y}}, Petr and {Moorhead},
        Althea V. and {Kuchner}, Marc J. and
        {Szalay}, Jamey R. and {Malaspina},
        David M.},
    title = "{How long-lived grains dominate the
        shape of the zodiacal cloud}",
    journal = {Planetary Science Journal},
    year = 2024,
    volume = {5},
    number = {3},
    eid = {82},
    pages = {82},
    doi = {10.3847/PSJ/ad2de8},
}

@article{poppe19,
    author = {{Poppe}, A.~R.},
    title = "{The contribution of centaur-emitted
        dust to the interplanetary dust
        distribution}",
    journal = {MNRAS},
    year = 2019,
    volume = {490},
    number = {2},
    pages = {2421-2429},
    doi = {10.1093/mnras/stz2800},
}

@article{sachse18,
    author = {{Sachse}, Manuel},
    title = "{A planetary dust ring generated by
        impact-ejection from the Galilean
        satellites}",
    journal = {Icarus},
    year = 2018,
    volume = {303},
    pages = {166-180},
    doi = {10.1016/j.icarus.2017.10.011},
}

@article{scipy,
  author  = {Virtanen, Pauli and Gommers, Ralf and Oliphant, Travis E. and
            Haberland, Matt and Reddy, Tyler and Cournapeau, David and
            Burovski, Evgeni and Peterson, Pearu and Weckesser, Warren and
            Bright, Jonathan and {van der Walt}, St{\'e}fan J. and
            Brett, Matthew and Wilson, Joshua and Millman, K. Jarrod and
            Mayorov, Nikolay and Nelson, Andrew R. J. and Jones, Eric and
            Kern, Robert and Larson, Eric and Carey, C J and
            Polat, {\.I}lhan and Feng, Yu and Moore, Eric W. and
            {VanderPlas}, Jake and Laxalde, Denis and Perktold, Josef and
            Cimrman, Robert and Henriksen, Ian and Quintero, E. A. and
            Harris, Charles R. and Archibald, Anne M. and
            Ribeiro, Ant{\^o}nio H. and Pedregosa, Fabian and
            {van Mulbregt}, Paul and {SciPy 1.0 Contributors}},
  title   = {{{SciPy} 1.0: Fundamental Algorithms for Scientific
            Computing in Python}},
  journal = {Nature Methods},
  year    = {2020},
  volume  = {17},
  pages   = {261--272},
  adsurl  = {https://rdcu.be/b08Wh},
  doi     = {10.1038/s41592-019-0686-2},
}

@article{sekii15,
    author = {{Sekii}, Takashi and {Appourchaux},
        Thierry and {Fleck}, Bernhard and
        {Turck-Chi{\`e}ze}, Sylvaine},
    year = 2015,
    title = "{Future mission concepts for
        helioseismology}",
    journal = {Space Science Reviews},
    volume = {196},
    number = {1-4},
    pages = {285-302},
    doi = {10.1007/s11214-015-0142-2},
}

@article{soja19,
    author = {{Soja}, R.~H. and {Gr{\"u}n}, E. and
        {Strub}, P. and {Sommer}, M. and
        {Millinger}, M. and {Vaubaillon}, J. and
        {Alius}, W. and {Camodeca}, G. and
        {Hein}, F. and {Laskar}, J. and
        {Gastineau}, M. and {Fienga}, A. and
        {Schwarzkopf}, G.~J. and {Herzog}, J.
        and {Gutsche}, K. and {Skuppin}, N. and
        {Srama}, R.},
    title = "{imem2: a meteoroid environment model
        for the inner solar system}",
    journal = {Astronomy and Astrophysics},
    year = 2019,
    volume = {628},
    eid = {A109},
    pages = {A109},
    doi = {10.1051/0004-6361/201834892},
}

@article{staubach97,
    author = {{Staubach}, P. and {Gr{\"u}n}, E. and
        {Jehn}, R.},
    title = "{The meteoroid environment near Earth}",
    journal = {Advances in Space Research},
    year = 1997,
    volume = {19},
    number = {2},
    pages = {301-308},
    doi = {10.1016/S0273-1177(97)00017-3},
}

@article{steel85,
    author = {{Steel}, D.~I. and {Baggaley}, W.~J.},
    title = "{Collisions in the solar system - I. Impacts of the Apollo-Amor-Aten asteroids upon the terrestrial planets}",
    journal = {MNRAS},
    year = 1985,
    volume = {212},
    pages = {817-836},
    doi = {10.1093/mnras/212.4.817},
}

@article{vokrouhlicky12,
    author = {{Vokrouhlick{\'y}}, David and
        {Pokorn{\'y}}, Petr and {Nesvorn{\'y}},
        David},
    year = 2012,
    title = "{{\"O}pik-type collision probability for
        high-inclination orbits}",
    journal = {Icarus},
    volume = {219},
    number = {1},
    pages = {150-160},
    doi = {10.1016/j.icarus.2012.02.021},
}

@inproceedings{weissman07,
    author = {{Weissman}, Paul R.},
    title = "{The cometary impactor flux at the
        earth}",
    booktitle = {Near Earth Objects, Our Celestial
        Neighbors: Opportunity and Risk},
    year = 2007,
    editor = {{Valsecchi}, G.~B. and
        {Vokrouhlick{\'y}}, D. and {Milani}, A.},
    series = {IAU Symposium},
    volume = {236},
    pages = {441-450},
    doi = {10.1017/S1743921307003559},
}

@article{west79,
    author = {D. H. D. West},
    year = {1979},
    title = {Updating mean and variance estimates},
    journal = {Communications of the {ACM}},
    volume = {22},
    number = {9},
    pages = {532-535},
    doi = {10.1145/359146.359153},
}

@article{wiegert09,
    author = {{Wiegert}, Paul and {Vaubaillon},
        Jeremie and {Campbell-Brown}, Margaret},
    title = "{A dynamical model of the sporadic
        meteoroid complex}",
    journal = {Icarus},
    year = 2009,
    volume = {201},
    number = {1},
    pages = {295-310},
    doi = {10.1016/j.icarus.2008.12.030},
}

\clearpage

\onecolumn

\appendix

\section{Spatial distribution}

\subsection{Derivation}
\label{sec:derivpdf}

We begin with two sets of assumptions: first, we assume that the semi-major axis ($a$), eccentricity ($e$), and inclination ($i$) of the meteoroid are known and do not change with time:
\begin{align}
    f_a(a') = \delta(a' - a) \, , ~\,~
    f_e(e') = \delta(e' - e) \, , ~ \text{ and } ~
    f_i(i') = \delta(i' - i) \, ,
\end{align}
where $\delta$ is the Dirac delta function.
Furthermore, we assume that all values of argument of pericenter ($\omega$), longitude of ascending node ($\Omega$), and mean anomaly ($M$) are equally likely; that is, that their PDFs are:
\begin{flalign}
    f_M(M) &= f_\omega(\omega) = f_\Omega(\Omega) = \frac{1}{2 \pi} \label{eq:pdf2pi}
\end{flalign}
We can then derive the PDF of the eccentric anomaly using the change-of-variable technique \citep[see, e.g.,][]{hogg20}:
\begin{align}
    f_E(E) &= f_M(M(E)) \times \left\lvert \frac{d M}{d E} \right\rvert
\end{align}
We know that $M = E - e \sin E$, and therefore $\tfrac{d M}{d E} = 1 - e \cos E$. We then obtain:
\begin{align}
    f_E(E) &= \frac{1 - e \cos E}{2 \pi}
\end{align}
We use the same process to derive the PDF of the true anomaly from that of eccentric anomaly. First, we express $f_E(E)$ in terms of $\nu$:
\begin{align}
    1 - e \cos E &= \frac{1 - e^2}{1 + e \cos \nu}
\end{align}
Next, we take the derivative of $E(\nu)$:
\begin{align}
    E &= 2 \tan^{-1} \left( \sqrt{\frac{1-e}{1+e}} \tan \frac{\nu}{2} \right) ~ \rightarrow ~
    \frac{d E}{d \nu} = \frac{\sqrt{1 - e^2}}{1+e \cos \nu}
\end{align}
This gives us the following PDF for $\nu$:
\begin{align}
    f_\nu(\nu) &= \frac{1}{2 \pi} \frac{(1-e^2)^{3/2}}{(1 + e \cos \nu)^2}
    \label{eq:fnu}
\end{align}

Consider the joint PDF of true anomaly and argument of perihelion. Because these two quantities are independent, their joint PDF is the product of their individual PDFs:
\begin{align}
    f_{\nu \omega} &= \frac{1}{4 \pi^2} 
    \frac{(1-e^2)^{3/2}}{(1 + e \cos \nu)^2}
    \label{eq:fnw}
\end{align}
We would like to transform this into a PDF for the meteoroid's location. In order to do so, we will define a dimensionless heliocentric distance, $s$, and a dimensionless ecliptic separation, $\xi$:
\begin{align}
    s &= (r-a)/ae = - \cos E \label{eq:sdef} \\
    \xi &= z/(r \sin i) = \sin(\omega + \nu) \label{eq:xidef}
\end{align}
In order to apply the change of variables technique, we must express $\nu$ and $\omega$ in terms of $s$ and $\xi$. We know that ${\xi = \sin(\omega + \nu)}$, and it can easily be shown -- using eq.~2.43 of \cite{murray99} -- that
\begin{align}
    \cos \nu &= - \frac{e + s}{1 - e s} \, .
\end{align}
This makes it simple to express $f_{\nu \omega}$ in terms of $s$ and $\xi$:
\begin{align}
    f_{\nu \omega}(\nu(s, \xi), \, \omega(s, \xi)) &= 
    \frac{1}{4 \pi^2} \frac{(1 + e s)^2}{\sqrt{1 - e^2}}
\end{align}

These equations have four solutions in total:
\begin{align}
    \nu_1 &= \cos^{-1} \left( - \frac{e + s}{1 - e s} \right) &
    \nu_2 &= \cos^{-1} \left( - \frac{e + s}{1 - e s} \right) &
    \nu_3 &= - \cos^{-1} \left( - \frac{e + s}{1 - e s} \right) &
    \nu_4 &= - \cos^{-1} \left( - \frac{e + s}{1 - e s} \right) 
        \label{eq:nuj} \\
    \omega_1 &= \sin^{-1} \xi - \nu_1 &
    \omega_2 &= \pi - \sin^{-1} \xi - \nu_2 & 
    \omega_3 &= \sin^{-1} \xi - \nu_3 & 
    \omega_4 &= \pi - \sin^{-1} \xi - \nu_4 
        \label{eq:wj}
\end{align}
We must therefore sum over all four solutions:
\begin{align}
    f_{s \xi}(s, \xi) &= \sum_j f_{\nu \omega}(\nu_j, \omega_j) \cdot \left\lvert \bm{\mathrm{J}}_{\nu_j \omega_j, \, s \xi} \right\rvert
\end{align}
where the last component is the determinant of the Jacobian matrix,
\begin{align}
    \bm{\mathrm{J}}_{\nu_j \omega_j, \, s \xi} &= \begin{pmatrix}
    \dfrac{\partial \nu_j}{\partial s} &
    \dfrac{\partial \nu_j}{\partial \xi} \\[1.2em]
    \dfrac{\partial \omega_j}{\partial s} &
    \dfrac{\partial \omega_j}{\partial \xi}
    \end{pmatrix}
    = \begin{pmatrix}
    \dfrac{\text{sgn}(\nu_j)}{1+e s}
        \sqrt{\dfrac{1-e^2}{1-s^2}} &
    0 \\[1.2em]
    - \dfrac{1}{1+e s}
        \sqrt{\dfrac{1-e^2}{1-s^2}} &
    \dfrac{\text{sgn}(\cos (\omega_j + \nu_j))}{\sqrt{1-\xi^2}} 
    \end{pmatrix}
\end{align}
We find that the absolute value of the determinant of this matrix is the same for all four solutions. Thus,
\begin{align}
    f_{s \xi}(s, \xi) &= 
    4 
    \times 
    \frac{1}{4 \pi^2} 
        \frac{(1 + e s)^2}{\sqrt{1 - e^2}} 
    \times
    \frac{1}{1+e s}
        \sqrt{\dfrac{1-e^2}{1-s^2}}
        \sqrt{\dfrac{1}{1 - \xi^2}} = 
    \frac{1}{\pi^2}
    \frac{1 + e s}{\sqrt{(1 - s^2)}}
    \frac{1}{\sqrt{1 - \xi^2}}
\end{align}

\subsection{Equivalence to previous works}
\label{sec:equiv}

Our dimensionless variables $s$ and $\xi$ are given in terms of $r$ and $z$ in equations\,\ref{eq:sdef} and \ref{eq:xidef}. The corresponding Jacobian is:
\begin{align}
    \bm{\mathrm{J}}_{s \xi, \, r z} &= \begin{pmatrix}
    \dfrac{\partial s}{\partial r} &
    \dfrac{\partial s}{\partial z} \\[1.2em]
    \dfrac{\partial \xi}{\partial r} &
    \dfrac{\partial \xi}{\partial z}
    \end{pmatrix}
    = \begin{pmatrix}
    \dfrac{1}{a e} &
    0 \\[1.2em]
    \dfrac{-z}{r^2 \sin i} &
    \dfrac{1}{r \sin i}
    \end{pmatrix}
\end{align}
Hence,
\begin{align}
    f_{r z}&(r, z) = f_{s \xi}(s(r, z), \xi(r, z)) \cdot \left\lvert \bm{\mathrm{J}}_{s \xi, \, r z} \right\rvert = \frac{r/a}{\pi^2} 
    \frac{1}{\sqrt{(a e)^2 - (r-a)^2}}
    \frac{1}{\sqrt{r^2 \sin^2 i - z^2}}
    \label{eq:frz}
\end{align}
is the joint PDF for heliocentric distance and ecliptic separation. Meteoroids with a given set of $a$, $e$, and $i$ values and evenly distributed $M$ and $\omega$ values will contribute to the local number density at $(r, z)$ in proportion to $f_{r z}(r, z)$. Note that $f_{rz}$ has singularities at $|r-a| = a e$ (where the radial velocity is zero) and also at $|z| = r \sin i$.

In order to obtain the number density, we must divide $f_{r z}$ by ${2 \pi r}$:
\begin{align}
    \eta &= \frac{1}{2 \pi r} f_{r z} (r, z) 
    = \frac{1}{2 \pi^3} \frac{1}{r a} \frac{1}{\sqrt{(a e)^2 - (r-a)^2}}
    \frac{1}{\sqrt{\sin^2 i - (z/r)^2}} 
\end{align}
If we substitute ${\sin \beta}$ for ${z/r}$, we recover eq.~21 of \cite{kessler81} and eq.~25 of \cite{steel85}.

\subsection{Volume of integration}
\label{sec:vol}

The volume bounded by ${r-d, r+d}$ and ${\beta - d/r, \beta + d/r}$ is:
\begin{align}
    V(r, \beta, d) &= 
    \int_0^{2 \pi} 
    \int_{\beta-d/r}^{\beta+d/r} 
    \int_{r-d}^{r+d} 
        (r')^2 \cos \beta' 
    \, d r' 
    \, d \beta' 
    \, d \lambda'
    = \frac{8 \pi}{3} d (d^2 + 3 r^2) \cos \beta \sin (d/r)
\end{align}
When $d$ is small compared to $r$, the volume is approximately
\begin{align}
    \lim_{d \rightarrow 0} V(r, \beta, d) = 8 \pi d^2 r \cos \beta
\end{align}

If ${\beta + d/r > \pi/2}$ or ${\beta - d/r < -\pi/2}$, the above integration limits are not meaningful. Instead, we must substitute $\pi/2$ for our upper integration limit or $-\pi/2$ for our lower integration limit. Then,
\begin{align}
    V(r, \beta, d) &= 
    \int_0^{2 \pi} \int_{\lvert \beta \rvert -d/r}^{\pi/2} \int_{r-d}^{r+d} (r')^2 \cos \beta' \, d r' \, d \beta' \, d \lambda'
    = \frac{4 \pi}{3} d (d^2 + 3 r^2) \left( 1 - \sin (\lvert \beta \rvert - d/r) \right)
\end{align}
When $d$ is small compared to $r$ (and therefore ${\pi - \lvert \beta \rvert - d/r}$ is also small), the volume is approximately
\begin{align}
    \lim_{d \rightarrow 0} V(r, \beta, d) = 2 \pi d r^2 \left( \frac{\pi}{2} - \lvert \beta \rvert + \frac{d}{r} \right)^2
\end{align}

We do not consider the case in which ${\beta + d/r > \pi/2}$ \emph{and} ${\beta - d/r < -\pi/2}$, because this requires the implausibly large value ${d > \pi r/2}$.


\section{Encounter geometry}

\subsection{Derivation}
\label{sec:derivvel}

Consider a spacecraft located at a given heliocentric distance $r$, whose distance from the ecliptic plane is ${z 
= r \sin \beta}$. Now suppose that this spacecraft encounters a meteoroid with orbital elements $a$, $e$, and $i$. 
Recall that we have defined ${s = (r-a)/a e = - \cos E}$ and ${\xi = z / r \sin i = \sin(\omega + \nu)}$. We can solve for $\nu$ and $\omega$ as follows
\begin{align}
    \cos \nu &= - \frac{e+s}{1+e s} 
        \label{eq:cf} \\
    \sin(\omega + \nu) &= \xi \, ,
\end{align}
where eq.\,\eqref{eq:cf} is derived from eq.~2.43 of \cite{murray99}. Note that each of these equations has two solutions.

Because the meteoroid nodes are randomly distributed, the environment should not depend on the ecliptic longitude of the spacecraft. We therefore rotate our coordinate system about the $z$-axis such that ${y = 0}$; we then use eq.~2.122 of \cite{murray99} to write:
\begin{align}
x/r &= \cos \Omega \cos (\omega + \nu) - \sin \Omega \sin (\omega + \nu) \cos i  
 = \cos \beta \\
y/r &= \sin \Omega \cos (\omega + \nu) + \cos \Omega \sin (\omega + \nu) \cos i 
 = 0 \, .
\end{align}
These equations can be re-arranged to obtain $\Omega$:
\begin{align}
    \sin \Omega &= - \frac{\sin (\omega + \nu)}{\cos \beta} \cos i \, , ~ \, ~
    \cos \Omega = \frac{\cos (\omega + \nu)}{\cos \beta}
\end{align}
Now that we have obtained the orbit angles $\nu$, $\omega$, and $\Omega$ that allow a meteoroid with a given value of $a$, $e$, and $i$ to intersect the location $x$, ${y=0}$, and $z$, we can calculate the corresponding meteoroid velocity vector. First, we express the velocity in terms of 
\begin{align}
    \vec{v} &= 
    \begin{pmatrix}
        \cos \Omega & - \sin \Omega & 0 \\
        \sin \Omega & \cos \Omega & 0 \\
        0 & 0 & 1
    \end{pmatrix} \cdot
    \begin{pmatrix}
        1 & 0 & 0 \\
        0 & \cos i & - \sin i \\
        0 & \sin i & \cos i
    \end{pmatrix} \cdot
    \begin{pmatrix}
        \cos \omega & - \sin \omega & 0 \\
        \sin \omega & \cos \omega & 0 \\
        0 & 0 & 1
    \end{pmatrix} \cdot u
    \begin{pmatrix}
        - \sin f \\
        e + \cos f \\
        0
    \end{pmatrix}
\end{align}
where 
\begin{align}
    u &= \sqrt{G M_\odot/(a(1-e^2))} \label{eq:v0} 
\end{align}
The above matrix multiplication yields
\begin{align}
    v_x &= - u \, e_s \cos \Omega - u \, e_c \sin \Omega \cos i  \\
    v_y &= - u \, e_s \sin \Omega + u \, e_c \cos \Omega \cos i  \\
    v_z &= + \, u \, e_c \sin i 
\end{align}
where 
\begin{align}
    e_s &= e \sin \omega + \sin (\omega + \nu) \label{eq:ap_es} \\
    e_c &= e \cos \omega + \cos (\omega + \nu) \label{eq:ap_ec} 
\end{align}
Once $\vec{v}$ has been determined, we can rotate it back into our inertial reference frame in which $y$ is not necessarily 0.

\end{document}